\documentstyle[sprocl]{article}
\input{psfig}
\def\Journal#1#2#3#4{{#1} {\bf #2}, #3 (#4)}

\def\PRD{{\em Phys. Rev.} D}
\def\ZPC{{\em Z. Phys.} C}

\def\be{\begin{equation}}
\def\ee{\end{equation}}
\def\bea{\begin{eqnarray}}
\def\eea{\end{eqnarray}}

\begin{document}
\begin{titlepage}
\hbox to \hsize{
\hfill\vtop{\hbox{}
\vspace{-2.cm}
\hbox{\large hep-ph/9609315}
\vspace{2.mm}
\hbox{\large UB-HET-96-02}
\vspace{2.mm}
\hbox{\large Fermilab-Conf-96/305-T}
\vspace{2.mm}
\hbox{\large August 1996} } }
\vspace{1.5cm}
\begin{center}
{\Large \sc Electroweak Radiative Corrections to $W$ and $Z$ Boson Production
in Hadronic Collisions\footnote{Talk given by U.B. at the
DPF96 Conference, Minneapolis, MN, August~10 --~15, 1996, to appear in
the Proceedings}\\[1.cm]}
{\large \sc U.~Baur\\[2.mm]}
{\large \it Physics Department, State University of New York at 
Buffalo, Buffalo, NY 14260, USA\\[0.5cm]}
{\large \sc S.~Keller\\[2.mm]}
{\large \it Fermilab, P.O. Box 500, Batavia, IL 60510, USA\\[0.5cm]}
{\large \sc W.~Sakumoto\\[2.mm]}
{\large \it Physics Department, University of Rochester, Rochester, NY
14627\\[0.5cm]}
{\large \sc D.~Wackeroth\\[2.mm]}
{\large \it Fermilab, P.O. Box 500, Batavia, IL 60510, USA\\[0.5cm]}
\vspace{1.cm}
{\large \bf
ABSTRACT \\[6.mm]}
{ \baselineskip=16pt \large
Some results of a calculation of electroweak radiative corrections to 
$W$ and $Z$ boson production in hadronic collisions are presented. 
}
\end{center} 
\end{titlepage}
\newpage

\title{ELECTROWEAK RADIATIVE CORRECTIONS TO $W$ AND $Z$ BOSON PRODUCTION
IN HADRONIC COLLISIONS}
\author{U.~BAUR }
\address{Physics Department, SUNY Buffalo, Buffalo, NY 14260}
\author{S.~KELLER, D.~WACKEROTH }
\address{Fermilab, P.O.~Box 500, Batavia, IL 60510}
\author{W.~SAKUMOTO}
\address{Physics Department, University of Rochester, Rochester, NY
14627}
\maketitle\abstracts{
Some results of a calculation of electroweak radiative corrections to 
$W$ and $Z$ boson production in hadronic collisions are presented. }

Over the last five years, the Standard Model of electroweak
interactions has been successfully tested at the one loop level.  
Experiments at LEP and the SLC have determined the properties of the $Z$
boson with a precision of 0.1\% or better, and the
range of the top quark mass has been correctly predicted by comparison with 
loop corrections.  Currently, the $Z$
boson mass is known to $\pm 2.2$~MeV, whereas the uncertainty of the $W$
boson mass, $M_W$, is $\pm 125$~MeV. A precise measurement of
$M_W$ and the top quark mass would make it possible to derive 
indirect constraints on the Higgs boson mass via top quark and
Higgs boson electroweak radiative corrections to $M_W$. 

A significant reduction of the $W$ boson mass uncertainty is expected 
from measurements at LEP2 and the upgraded Tevatron. The
ultimate precision expected for $M_W$ from the LEP2 experiments combined is 
approximately 40~MeV.  At the
Tevatron, integrated luminosities of ${\cal O}(1~{\rm fb}^{-1})$ are 
envisioned in the Main Injector Era, resulting in an
uncertainty of approximately 50~MeV per experiment.  
Further upgrade of the Tevatron might be possible with a goal of
an overall integrated luminosity of ${\cal O}$(30~fb$^{-1}$) and a 
precision on $M_W$ of about 15~MeV~\cite{TEV33}.

The determination of the $W$ boson mass in a hadron collider environment
requires a simultaneous precision measurement of $M_Z$.
When compared to the value measured at LEP,
the $Z$ boson mass helps to accurately determine the energy scale and
the resolution of the charged lepton energy.

In order to measure $M_W$ and $M_Z$ with high precision, it is crucial 
to fully control higher order QCD and electroweak (EW) corrections. 
In this short contribution, we concentrate on the latter. So far, only 
the final state photonic corrections have been calculated~\cite{BK}, using an 
approximation which indirectly estimates the virtual part
from the inclusive ${\cal O}(\alpha^2)$ $W\to\ell\nu(\gamma)$
and $Z\to\ell^+\ell^-(\gamma)$ width and the hard photon bremsstrahlung 
contribution.  Here we summarize the results of a calculation 
which includes both initial and (exact) final state EW corrections, 
as well as initial -- final state interference~terms. 

Our calculation is performed using standard Monte Carlo phase space
slicing techniques for
NLO calculations. The resulting $p\,p\hskip-7pt\hbox{$^{^{(\!-\!)
}}$} \rightarrow \gamma^*,\, Z \rightarrow \ell^+ \ell^-(\gamma)$ and 
$p\,p\hskip-7pt\hbox{$^{^{(\!-\!) }}$} \rightarrow W\to\ell\nu(\gamma)$ 
cross sections are independent of the soft and collinear cutoff
parameters used to divide the phase space into $2\to 2$ and $2\to 3$
regions.  Special care has to be taken in calculating the initial state 
radiative corrections: mass (collinear) singularities have 
to be absorbed into the parton distribution functions (PDF) through 
factorization, in complete analogy to the QCD case. However, EW 
corrections to the PDF evolution are not included in our calculation; 
they are expected to be small~\cite{SKP}. For $Z$ boson production, 
purely weak corrections are very small and therefore are not included. In
the $W$ case, weak corrections were included in the calculation 
as required by gauge invariance, and the corrections were 
consistently separated into
gauge invariant initial and final state contributions~\cite{WH}. The 
technical details of our calculations will be presented in 
Refs.~\ref{REF:BKW} and~\ref{REF:BKS}.
\begin{figure} 
\begin{center}
\begin{tabular}{cc}
\psfig{figure=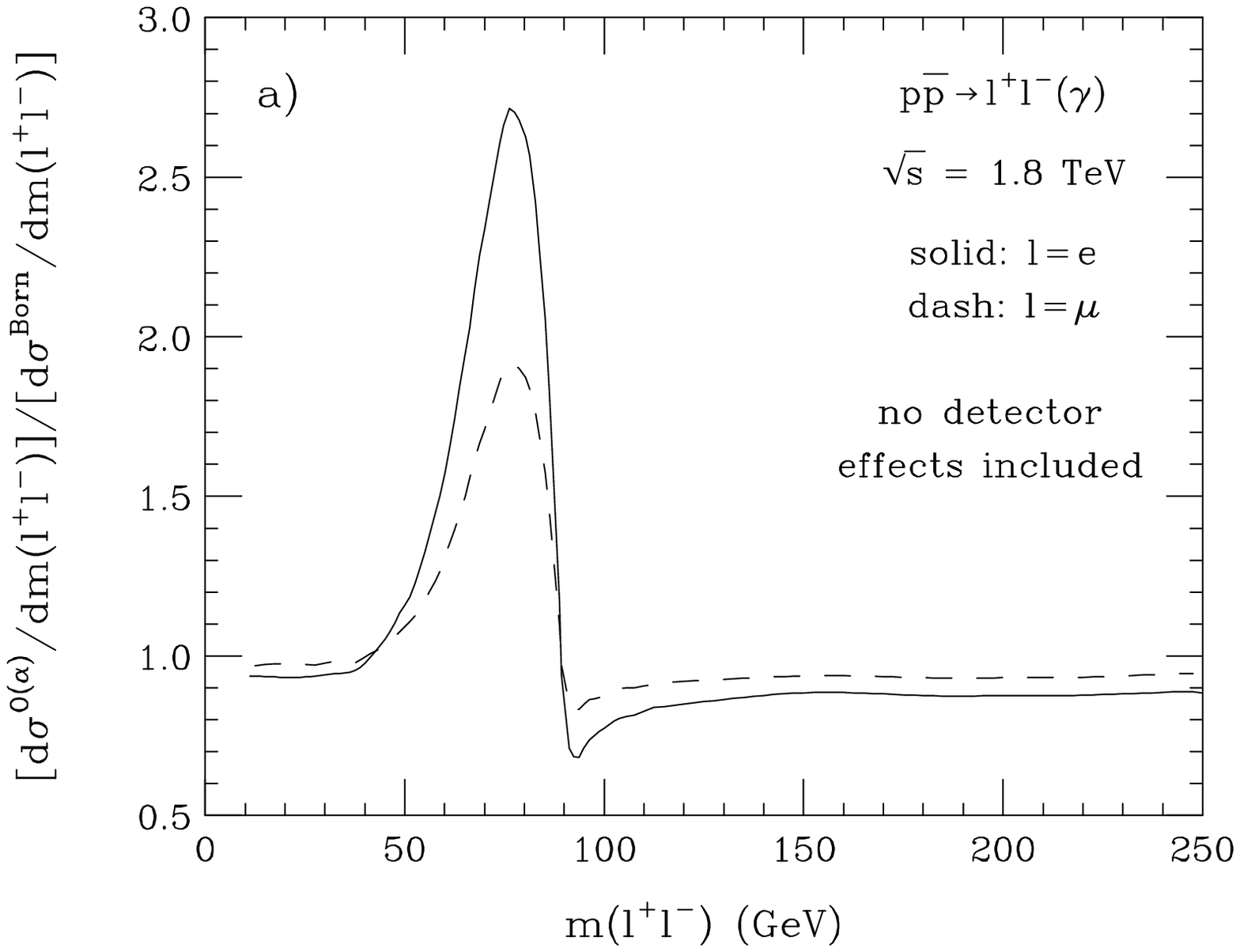,height=1.65in} &
\psfig{figure=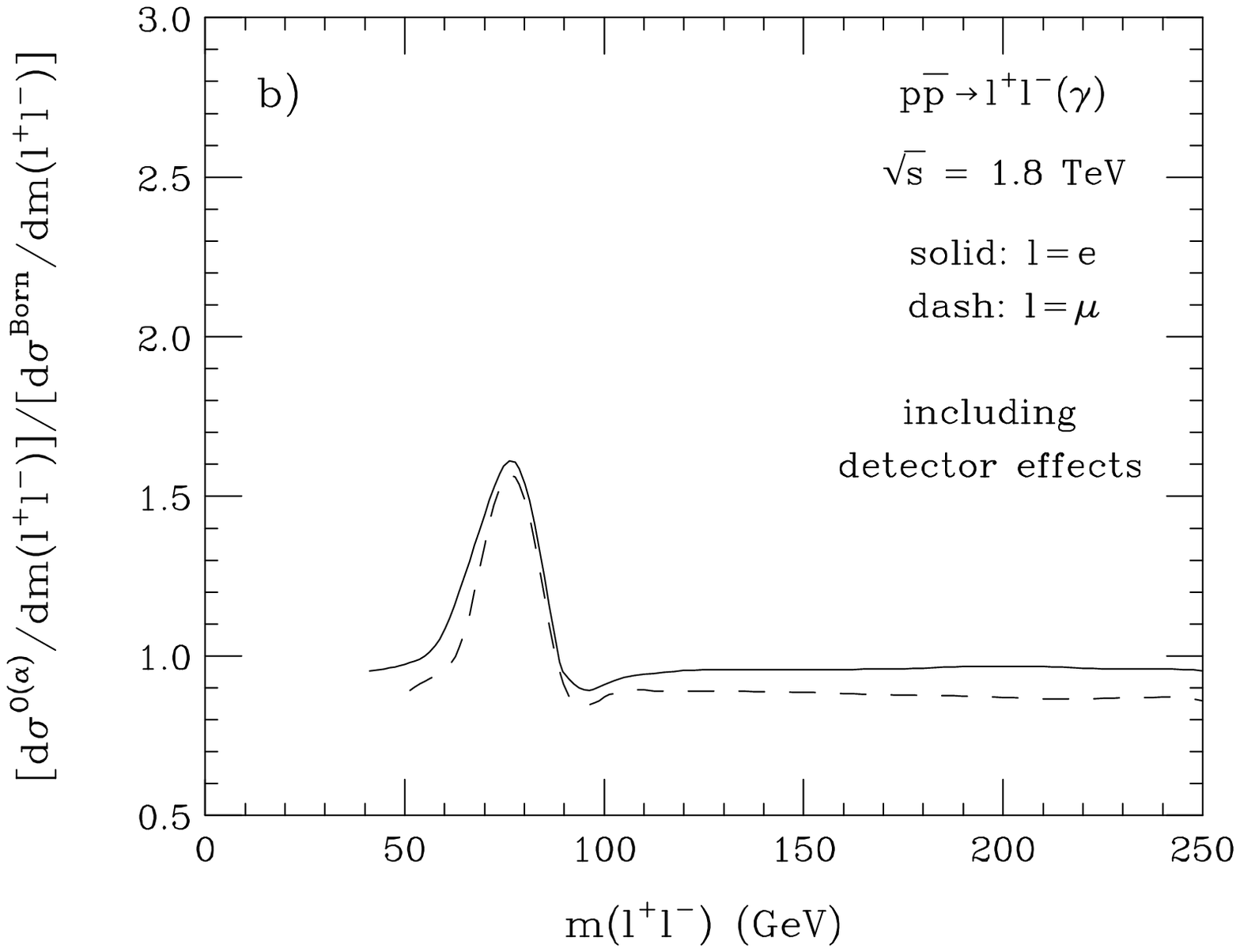,height=1.65in} 
\end{tabular}
\caption{Ratio of NLO to LO differential cross sections for $Z$ boson 
production as a function of the di-lepton invariant mass at the Tevatron.}
\label{FIG:ONE}
\end{center}
\end{figure}

As expected, we find that the final state corrections completely
dominate over the entire mass range of interest. 
Both the initial state and the interference contributions are small 
and therefore have a negligible impact on the extracted $W/Z$ boson mass.
In Fig.~\ref{FIG:ONE}, we show the ratio of NLO to LO differential cross
sections for $Z$ boson production as a 
function of the di-lepton invariant mass, $m(\ell^+\ell^-)$.  
When no detector effects are included (Fig.~\ref{FIG:ONE}a), the corrections
are large due to the occurrence of mass singular logarithms
$\log(M_Z^2/m_\ell^2)$ ($m_\ell$ denotes the lepton mass)
in the collinear region. These logarithms also explain the large difference 
in the differential cross section between electrons and muons.
The collinear behavior of the final state corrections also 
significantly influences the shape of the
$m(\ell^+\ell^-)$ distribution, as a large fraction of the events
shifts from the peak region to lower values of $m(\ell^+ \ell^-)$.
In Fig.~\ref{FIG:ONE}b, the same ratio is plotted as in 
Fig.~\ref{FIG:ONE}a, now taking into account CDF detector resolution
effects. While the mass singular logarithms are removed by the detector 
effects in the electron case, residual effects of the
$\log(M_Z^2/m_\ell^2)$ terms remain in the muon case due to differences
in the experimental criteria used to identify electrons and
muons~\cite{BKS}.

EW corrections have a smaller effect on the $W$ transverse mass distribution 
than on the $Z$ invariant mass distribution; 
including detector effects, they reduce the cross section by a few per
cent in the peak region. The shifts in the $W/Z$ boson mass extracted from
the transverse/invariant mass distribution that we obtained are
consistent  with the values obtained by the CDF and D\O\ collaborations.  
A preliminary comparison of the approximation used so far~\cite{BK} with
our complete calculation reveals that the difference corresponds
to a shift in the mass of the order of 10 MeV.  

In summary, we have presented selected results of a calculation of 
weak boson production in hadronic collisions which includes 
initial and final state EW corrections, as well as initial--final 
state interference terms.  Final state
corrections dominate, while initial state corrections are small after
factorizing the collinear singularities into the parton distribution 
functions. Our calculation substantially improves the treatment of 
EW corrections to $W/Z$ production at hadron colliders, and will allow
to considerably reduce the systematic uncertainties associated with the 
EW corrections when the $W$ and $Z$ boson masses are extracted from 
Tevatron data.

\end{document}